# Cryogenic temperature deposition of high-performance CoFeB/MgO/CoFeB magnetic tunnel junctions on ϕ300 mm wafers

Tomohiro Ichinose,[1*] Tatsuya Yamamoto,[1] Takayuki Nozaki,[1] Kay Yakushiji,[1] Shingo Tamaru,[1] Makoto Konoto,[1] Shinji Yuasa[1]

1. National Institute of Advanced Industrial Science and Technology (AIST), Research Center for Emerging Computing Technologies, Tsukuba, Japan.

*Corresponding author

e-mail: tomohiro.ichinose@aist.go.jp

*Abstract*

We developed a cryogenic temperature deposition process for high-performance CoFeB/MgO/CoFeB magnetic tunnel junctions (MTJs) on ϕ300 mm thermally oxidized silicon wafers. The effect of the deposition temperature of the CoFeB layers on the nanostructure, magnetic and magneto-transport properties of the MTJs were investigated in detail. When CoFeB was deposited at 100 K, the MTJs exhibited a perpendicular magnetic anisotropy (PMA) of 214 µJ/m$^2$ and a voltage-controlled magnetic anisotropy (VCMA) coefficient of -45 fJ/Vm, corresponding to 1.4- and 1.7-fold enhancements in PMA and VCMA, respectively, compared to the case of room-temperature deposition of CoFeB. The improvement in the MTJ properties was not simply due to the morphology of the MTJ films. The interface-sensitive magneto-transport properties indicated that interfacial qualities such as intermixing and oxidation states at the MgO/CoFeB interfaces were improved by the cryogenic temperature deposition. Cryogenic-temperature sputtering deposition is expected to be a standard manufacturing process for next-generation magnetoresistive random-access memory.

*Introduction*

Magnetoresistive random-access memory (MRAM) is a nonvolatile memory that uses magnetic tunnel junctions (MTJs) to store information. Among the various types of MRAM, spin-transfer-torque MRAM (STT-MRAM), which consists of MgO-based MTJs [1-4], has features such as relatively high read/write speed and relatively low write power consumption and is currently used as embedded memory in system LSI. On the other hand, voltage-controlled MRAM (VC-MRAM) is a candidate for next-generation embedded MRAM because it can more drastically reduce write power consumption compared with STT-MRAM [5-8]. The write operation of VC-MRAM is based on the voltage-induced dynamic switching and voltage-controlled magnetic anisotropy (VCMA) effect; its writing energy can be as small as approximately 1 fJ/bit with an operating speed in the GHz regime [5,6,9-17]. MTJs for STT-MRAM and VC-MRAM should have a high tunnel magnetoresistance (TMR) ratio and large perpendicular magnetic anisotropy (PMA). Moreover, VC-MRAM requires the MTJs to have a large VCMA coefficient. To achieve a large PMA and VCMA coefficient, the CoFeB-free layer (storage layer) of the MTJ needs to be ultrathin because both the interfacial PMA and VCMA effects originate solely from the barrier/free layer interface. Therefore, fabricating an ultrathin CoFeB layer with a high-quality interface is essential.

The standard MTJ structure for STT-MRAM and VC-MRAM is CoFeB/MgO/CoFeB with bottom-pinned and top-free CoFeB layers because this tri-layer structure can be fabricated on (111)-textured Co/Pt multilayers with a very high PMA [18], which gives rise to the pinning force of the bottom-pinned CoFeB layer [8,19-21]. However, the top CoFeB layer of this standard structure tends to have a low quality because island-like initial growth of CoFeB occurs on the MgO(001) surface, reflecting the poor wettability of CoFeB on MgO(001) caused by the low surface energy of MgO(001) [22-27]. The island-like initial growth results in a rougher CoFeB layer, which degrades magnetic properties such as the TMR ratio and PMA, particularly when the CoFeB layer is ultrathin [24,25,27].

Therefore, developing deposition processes to achieve a high-quality ultrathin CoFeB layer on MgO is of considerable importance for next-generation STT-MRAM and VC-MRAM. The quality of the CoFeB/MgO interface is also very important because PMA, TMR, and VCMA effects are interface-sensitive phenomena [28-32].

In this study, the effect of low-temperature deposition on MTJ properties was investigated. In most previous studies, CoFeB/MgO/CoFeB MTJ films were deposited at room temperature (RT) by sputtering; there have been very few reports on low-temperature sputtering deposition. Low-temperature deposition is expected to have a substantial influence on MTJ quality because it can suppress the migration of atoms on the film surface, resulting in the suppression of island-like growth as well as the suppression of intermixing and over-oxidation at the CoFeB/MgO interface. Note that suppressing the over-oxidation at the CoFeB/MgO interface effectively improves MTJ properties, according to first-principles calculations [30,33,34]. Therefore, low-temperature deposition is expected to have positive effects on PMA, TMR ratio, and VCMA coefficient. In this study, MTJ films were deposited at a cryogenic temperature (100 K) as well as at RT and the effect of the deposition temperature on the nanostructure, PMA, TMR ratio, and VCMA coefficient was investigated.

*Experimental Methods*

The top-free-type MTJ stacking structure of Ta(5)/Ru(5)/Ta(5)/Ru(5)/Ta(5)/CoFeB(3)/ MgO(2)/CoFeB($t$)/Mo(1)/Ta(3)/Ru(10) (thickness in nanometers) was deposited on $\phi$300 mm thermally oxidized Si substrates using a manufacturing-type sputtering system (EXIM, Tokyo Electron Ltd.). The sputtering system had seven sputtering deposition chambers with a base pressure of $3\times10^{-6}$ Pa. One of the deposition chambers was equipped with a low-temperature wafer stage cooled using a He refrigerator. The stage temperature was stabilized at 100–300 K. A $\phi$300 mm wafer was electrostatically chucked on the cooling stage and rotated during sputtering deposition, which enabled rapid and homogeneous cooling. The bottom and top CoFeB layers were deposited at a growth temperature $T_{\text{CoFeB}}$ of either 100 or 300 K. The remaining layers were deposited at a temperature of 300 K. The MgO barrier layer was prepared using repetitive sequences of metallic Mg deposition, followed by post-oxidation with an oxygen gas flow. *Ex-situ* thermal annealing at 573 K was performed in a vacuum furnace at $< 8 \times 10^{-5}$ Pa without a magnetic field.

The MTJs in this study have an orthogonal magnetization configuration under a zero magnetic field because the bottom and top CoFeB layers show in-plane and perpendicular magnetic anisotropy, respectively. It should be noted that the orthogonal configuration was intentionally employed to quantitatively evaluate the PMA and VCMA of the top CoFeB free layer from the magnetization and magnetoresistance curves, as discussed later. Note also that for MRAM applications, a strong perpendicular magnetic anisotropy can be given to the bottom-pinned layer by inserting the Co/Pt multilayer under the bottom CoFeB layer. The samples were patterned into elliptical shaped pillars of dimension $1.84 \times 0.84$ μm$^2$ using photolithography and Ar ion milling techniques. The magnetic properties of the MTJs were evaluated using vibrating-sample magnetometry (VSM). The structural properties were characterized using scanning transmission electron microscopy (STEM) and energy-dispersive X-ray spectroscopy (EDX). The magneto-transport properties of the MTJs were measured using a prober system with an electromagnet generating a magnetic field of up to 1 T.

*Experimental Results*

Figure 1 shows the magnetization curves of the unpatterned MTJ films for an in-plane applied field. The abrupt change in magnetization around zero magnetic fields in Figures. 1(a) and 1(c) is due to magnetization switching of the 3-nm-thick bottom CoFeB layer. On the other hand, the gradual increase in the magnetization in Figures 1(b) and 1(d) is due to the rotation of the magnetization of the top CoFeB layer from the perpendicular direction. The magnetization curves in Figure 1(a) ($T_{\text{CoFeB}} = 300$ K) have a rounded shape, which indicates a gradual saturation of magnetization compared with those in Figure 1(c) ($T_{\text{CoFeB}} = 100$ K). This means that the CoFeB layers deposited at 300 K have more spatial inhomogeneity in magnetic properties such as PMA than those deposited at 100 K. This indicates that the CoFeB layers deposited at 300 K were rougher and/or more intermixed than those deposited at 100 K. Namely, it is considered that the deposition at 100 K suppressed the intermixing and/or island-like growth of CoFeB on MgO(001), resulting in more homogeneous CoFeB layers. Figure 1(e) shows the thickness dependence of the saturation magnetization ($M_S \cdot t$) of the CoFeB layer deposited on the MgO(001) layer. For this measurement, control samples without the bottom CoFeB layer were used to precisely evaluate the saturation magnetization. The 100-K-deposited CoFeB showed a slightly larger magnetization compared with the 300 K-deposited CoFeB. This difference is mainly due to the difference in the thickness of the magnetic dead layer, which is 0.60 and 0.56 nm for $T_{CoFeB}$ = 300 and 100 K, respectively. This suggests that the 300-K-deposited CoFeB was more intermixed with MgO at the interface compared with the 100-K-deposited CoFeB.

    The nanostructures of the MTJs were analyzed using STEM, as shown in Figure 2. Figures 2(a) and 2(h) show cross-sectional bright-field STEM images of the MTJs deposited at $T_{CoFeB}$ = 300 and 100 K, respectively. Figures 2(b) and 2(i) show the high angle annular dark-field STEM (HAADF-STEM) images for the MTJs deposited at $T_{CoFeB}$ = 300 and 100 K, respectively. Figures 2(d)–(g) and 2(k)–(n) show EDX mapping images for Mg, Fe, Mo, and Ta. In the bright-field images in Figures 2(a) and 2(h), lattice patterns of the CoFeB/MgO/CoFeB structure were identified for both samples, indicating that the CoFeB layers crystallized from the interfaces with MgO(001) during post-annealing at 573 K [4]. Although the interface between the top CoFeB and Mo layers is not very clear in the bright- and dark-field STEM images, the separation of the CoFeB and Mo layers is clearly observed in the EDX mapping images for Fe and Mo in Figure 2. The EDX mapping images show no significant intermixing between layers; this may be because the spatial resolution of EDX mapping is not sufficiently high to identify atomic-scale intermixing and over-oxidation at the CoFeB/MgO interfaces.

For the quantitative evaluation of interfacial roughness, statistical analyses were performed for HAADF-STEM images of the CoFeB/MgO/CoFeB MTJ films, as shown in Figures 2(b) and 2(i). The broken lines in Figures 2(b) and 2(i) show the interfaces of the bottom CoFeB/MgO, MgO/top CoFeB, and Mo/Ta, which were obtained by applying the Hamiltonian Monte Carlo method to the modeled structures with three normally distributed interfaces. As mentioned previously, the top CoFeB/Mo interface could not be identified in the images. The estimated root-mean-square (RMS) values of the interfacial roughness are listed in Table 1. The roughness of each layer was basically the same for the 300-K- and 100-K-deposited samples within statistical errors. Therefore, the film morphology is not the main reason for the differences in magnetic properties between the 300-K- and 100-K-deposited MTJs.

Figure 3 shows the magneto-transport properties of the MTJs deposited at $T_{\text{CoFeB}}$ = 300 and 100 K. Figure 3(a) shows typical TMR curves measured at 50 mV by applying in-plane magnetic fields of up to 1 T along the long direction of the elliptic MTJs. The TMR ratio was defined as $\Delta R(H)/R_{min} \times 100$ (%), where $\Delta R(H) = R(H) - R_{min}$. $R(H)$ is the junction resistance as a function of the magnetic field $H$, and $R_{min}$ is the lowest junction resistance, which corresponds to the parallel magnetization configuration. Because the magnetization configuration is orthogonal at zero magnetic field, the highest TMR ratio, $\Delta R(0)/R_{min} \times 100$ (%), is half of the TMR ratio for the usual definition. The improved TMR ratio and PMA of the MTJs deposited at $T_{\text{CoFeB}}$ = 100 K are clearly observed in Figure 3(a). Figures 3(d) and 3(e) show the CoFeB-thickness-dependence of the TMR ratio and resistance-area ($RA$) product. The TMR ratio monotonically increases with the CoFeB thickness both for $T_{\text{CoFeB}}$ = 300 and 100 K. The MTJs deposited at 100 K show higher TMR ratios and lower $RA$ products compared with those deposited at 300 K for the whole thickness range. Figures 3(b) and 3(c) are the field dependence of normalized conductance $G$ derived from $(R(0) - R(H))/R(H) \times R_{min}/(R(0) - R_{min})$, which reflects the magnetization process of the top CoFeB [35].

As observed in the magnetization curves (Figure 1), sharper magnetization process was observed in the MTJs deposited at $T_{CoFeB}$ = 100 K. PMA energy per unit junction area, $K_u \cdot t$, was estimated by integrating the green shaded area in Figures 3(b) and 3(c) and by using the saturation magnetization values obtained from the VSM measurements. $K_u \cdot t$ is plotted as a function of the top CoFeB thickness in Figure 3(f). The PMA energy at the CoFeB thickness of 0.9 nm is 149 µJ/m² and 214 µJ/m² for MTJs deposited at $T_{CoFeB}$ = 300 K and 100 K, respectively. The MTJs deposited at $T_{CoFeB}$ = 100 K showed a higher PMA energy for all CoFeB thicknesses.

Figure 4(a) shows the bias-voltage dependence of the TMR ratios normalized at 50 mV. The bias direction was defined with respect to the top CoFeB electrode, as shown in the inset of Figure 4(a). For positive bias, the MTJs deposited at $T_{CoFeB}$ = 300 and 100 K showed nearly the same bias dependence of TMR. Conversely, for negative bias, the MTJs deposited at $T_{CoFeB}$ = 300 K exhibited a larger decrease in the TMR ratio compared with the MTJs deposited at $T_{CoFeB}$ = 100 K. Tunneling currents tend to strongly reflect the quality of the interface receiving the tunneling electrons [36]. Therefore, a larger bias dependence of the TMR indicates more scattering of tunneling electrons at the downstream interface. For a negative bias, electrons flow from the bottom to the top electrode. The larger bias dependence of the TMR observed for the MTJs deposited at $T_{CoFeB}$ = 300 K indicates that the MgO/top CoFeB interface for $T_{CoFeB}$ = 300 K is more defective compared with that for $T_{CoFeB}$ = 100 K.

Figure 4(b) shows the bias dependence of the PMA energy of the top CoFeB layer; the slope corresponds to the VCMA coefficient. The VCMA coefficient is plotted as a function of the CoFeB thickness, as shown in Figure 4(c). The VCMA coefficient increases with increasing CoFeB thickness. For all thickness ranges, the MTJs deposited at $T_{CoFeB}$ = 100 K showed a larger VCMA coefficient compared with the MTJs deposited at $T_{CoFeB}$ = 300 K. The largest VCMA coefficients were -26 and -45 fJ/Vm for $T_{CoFeB}$ = 300 and 100 K, respectively. The larger VCMA coefficient may reflect a better interface quality for $T_{CoFeB}$ = 100 K.

To summarize, the study showed differences in the properties of the MTJ fabricated at different deposition temperatures. The MTJs deposited at $T_{CoFeB}$ = 100 K exhibited (i) sharper magnetization saturation, (ii) thinner magnetic dead layer of the top CoFeB layer, (iii) higher TMR ratio, (iv) lower $RA$ product, (v) larger PMA, and (vi) larger VCMA coefficient, compared with the results of the MTJs deposited at $T_{CoFeB}$ = 300 K. These experimental results suggest that the MgO/top CoFeB interface of the MTJs deposited at $T_{CoFeB}$ = 100 K is less intermixed and less over-oxidized compared with the interface of those deposited at $T_{CoFeB}$ = 300 K. It should be noted that intermixing and over-oxidation at the interface prevent the coherent tunneling of $\Delta_1$ Bloch states, resulting in a reduced TMR ratio and higher $RA$ product. Moreover, the chemically clean interface, where Fe(Co)–O bonding occurs without excess oxygen atoms, showed the highest interfacial PMA. In addition to improvements in the TMR ratio and interfacial PMA, the large VCMA coefficient may have originated from the less intermixing high-quality interface in the MTJs deposited at $T_{CoFeB}$ = 100 K.

*Conclusion*

In summary, a cryogenic temperature deposition process was developed for the deposition of high-performance CoFeB/MgO/CoFeB MTJs on ϕ300 mm thermally oxidized silicon wafers and the effect of the deposition temperature on the nanostructure, magnetic, and magneto-transport properties of ultrathin CoFeB films were investigated. The nanostructure analyses using STEM indicated that the interfacial roughness in CoFeB/MgO/CoFeB was unchanged for the films deposited at 300 and 100 K. On the other hand, the magnetic and magneto-transport measurements revealed that the cryogenic temperature deposition clearly enhanced the PMA, TMR ratio, and VCMA coefficient. These enhancements can be attributed to the improvement in the interfacial qualities, such as less-intermixed and less-over-oxidized states at the MgO/top CoFeB interface resulting from the deposition at cryogenic temperatures. Cryogenic-temperature deposition is expected to be an effective manufacturing process for next-generation MRAM.

*Acknowledgments*

This study was funded in part by the New Energy and Industrial Technology Development Organization's (NEDO) Project (No. JPNP20017). The TEM analyses were performed by the Foundation for Promotion of Material Science and Technology of Japan (MST).

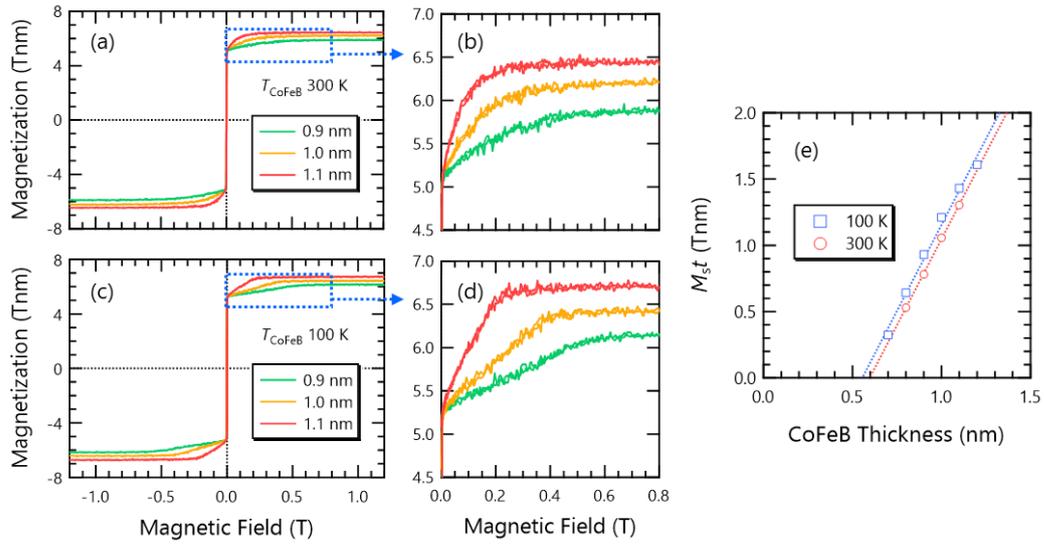

Figure 1. Magnetization curves of magnetic tunnel junctions (MTJs) consisting of CoFeB layers deposited at (a) 300 K and (c) 100 K. The blue rectangle regions in (a) and (c) are magnified as (b) and (d), respectively. (e) The top CoFeB layer thickness ($t$) dependence of $M_s t$, where $M_s$ is the saturation magnetization of top CoFeB.

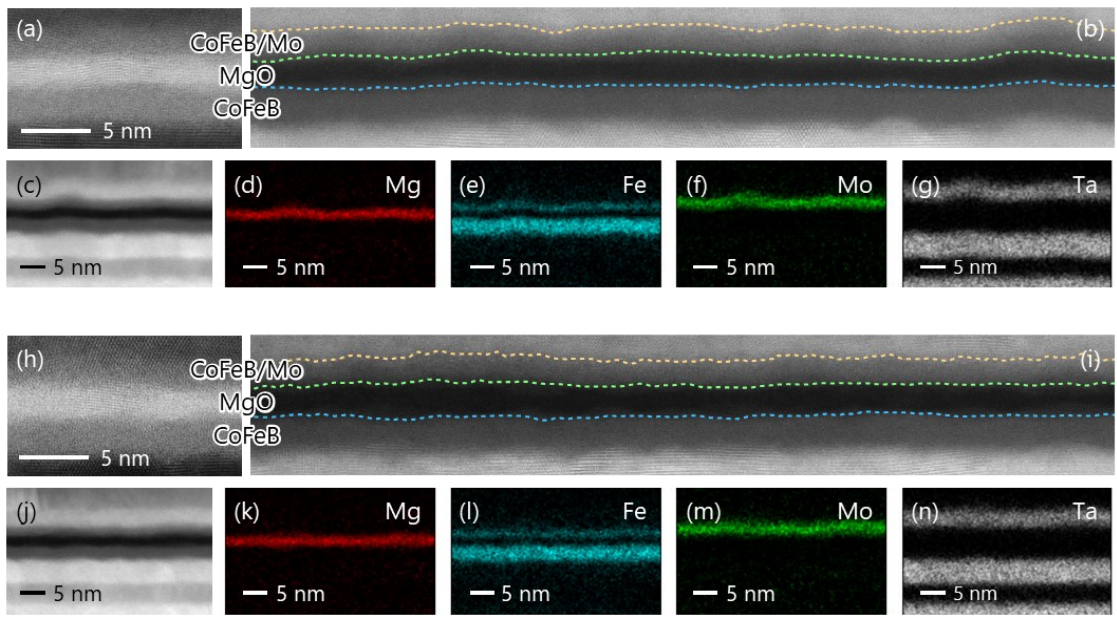

Figure 2. (a) and (b) [(h) and (i)] show the local and wide-area cross-sectional scanning transmission electron microscopy (STEM) images of the MTJs consisting of CoFeB layers deposited at 300 K [100 K]. (c)–(g) [(j)–(n)] High-angle annular dark field (HAADF)-STEM images and energy dispersive x-ray spectroscopy (EDX) maps for Mg, Fe, Mo, and Ta in MTJs consisting of CoFeB layers deposited at 300 K [100 K].

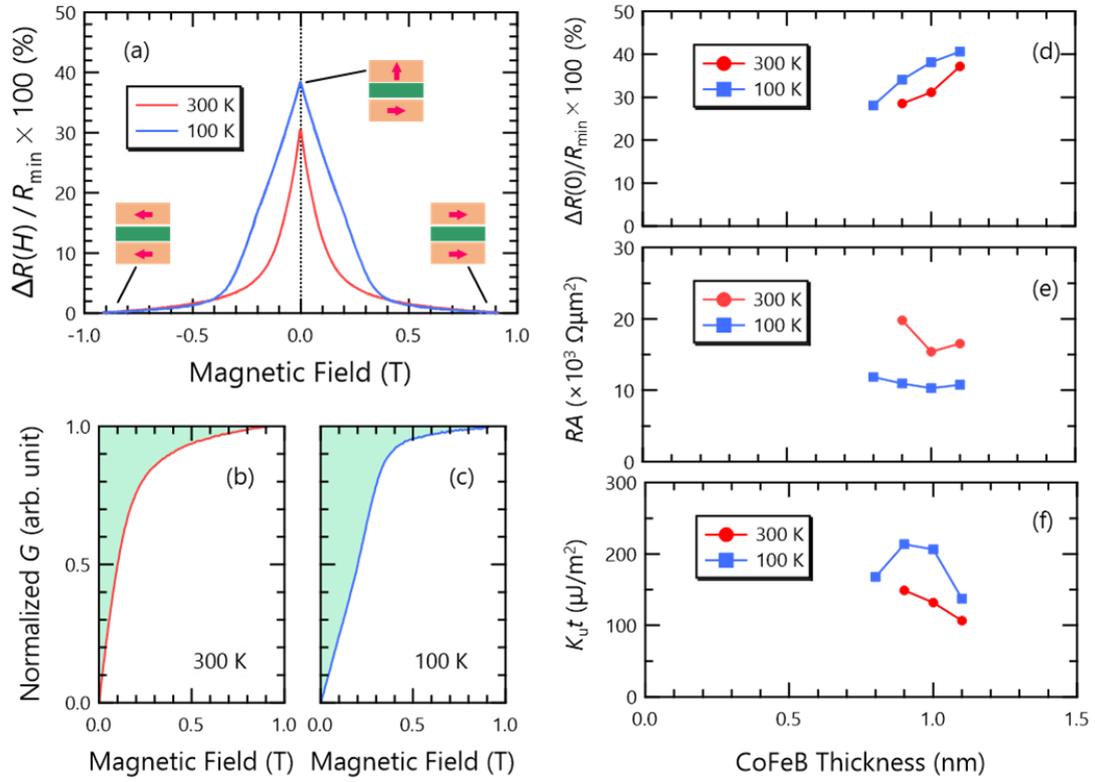

Figure 3. (a) Tunnel magnetoresistance (TMR) curves under in-plane magnetic field. (b) and (c) Normalized conductance ($G$) curves of MTJs consisting of CoFeB layers deposited at 300 and 100 K as a function of the magnetic field. (d) Half TMR ratios, (e) resistance-area ($RA$) products, and (f) perpendicular magnetic anisotropy (PMA) energy per unit junction area ($K_u t$) as a function of top CoFeB thickness.

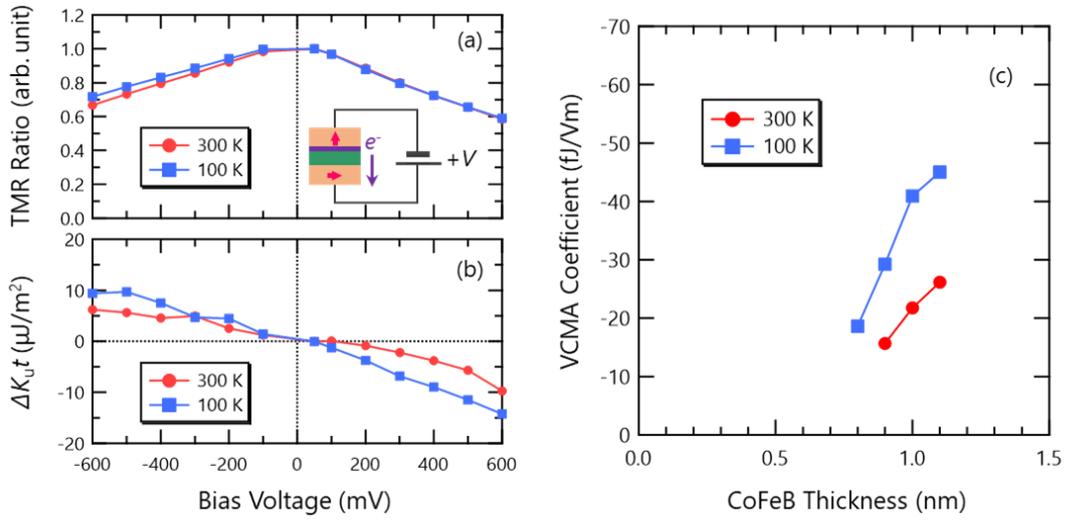

Figure 4. (a) Normalized TMR ratios and (b) variation of $K_u t$ as a function of bias voltage. The inset in (a) illustrates the bias voltage direction. (c) voltage controlled magnetic anisotropy (VCMA) coefficient as a function of the top CoFeB thickness

Table 1. Root-mean-square (RMS) values of the interfacial roughness of MTJs fabricated via the deposition of CoFeB layers at 300 and 100 K

| Interfaces | RMS roughness (nm) | |
| :---: | :---: | :---: |
| | 300 K | 100 K |
| Mo/Ta | 0.24±0.07 | 0.23±0.10 |
| MgO/CoFeB | 0.20±0.05 | 0.16±0.05 |
| CoFeB/MgO | 0.12±0.04 | 0.14±0.05 |